\documentclass[pra,twocolumn,superscriptaddress,aps,showpacs,amsmath,amssymb,nofootinbib]{revtex4}
\usepackage{graphicx}
\usepackage{dcolumn}
\usepackage{bm}
\usepackage{color}

\usepackage[section]{placeins}
\usepackage{graphicx,epsfig}
\usepackage{amsmath}
\usepackage{amsfonts}
\usepackage{braket}
\usepackage{fancyhdr}

\newcommand{\bce}{\begin{center}}
\newcommand{\ece}{\end{center}}
\newcommand{\beq}{\begin{equation}}
\newcommand{\eeq}{\end{equation}}
\newcommand{\beqa}{\begin{eqnarray}}
\newcommand{\eeqa}{\end{eqnarray}}

\begin{document}

\title{Quantum properties of a binary bosonic mixture in a double well}
\date{\today}
\author{Pere Mujal}
\affiliation{Departament d'Estructura i Constituents de la Mat\`{e}ria,\\
Universitat de Barcelona, 08028 Barcelona, Spain}
\author{Bruno Juli\'{a}-D\'{i}az}
\affiliation{Departament d'Estructura i Constituents de la Mat\`{e}ria,\\
Universitat de Barcelona, 08028 Barcelona, Spain}
\affiliation{Institut de Ci\`{e}ncies del Cosmos, Universitat de Barcelona, IEEC-UB, Mart\'{i} i Franqu\`{e}s 1, E-08028 Barcelona, Spain}
\affiliation{Institut de Ci\`{e}ncies Fot\`{o}niques, Parc Mediterrani de la Tecnologia, 08860 Barcelona, Spain}
\author{Artur Polls}
\affiliation{Departament d'Estructura i Constituents de la Mat\`{e}ria,\\
Universitat de Barcelona, 08028 Barcelona, Spain}
\affiliation{Institut de Ci\`{e}ncies del Cosmos, Universitat de Barcelona, IEEC-UB, Mart\'{i} i Franqu\`{e}s 1, E-08028 Barcelona, Spain}
\begin{abstract}
This work contains a detailed analysis of the properties of the ground state of a two-component two-sites Bose-Hubbard model, which captures the physics of a binary mixture of Bose-Einstein condensates trapped in a double-well potential.  The atom-atom interactions within each species and among the two species are taken as variable parameters while the hopping terms are kept fixed. To characterize the ground state we use observables such as the imbalance of population and its quantum uncertainty. The quantum many-body correlations present in the system are further quantified by studying the degree of condensation of each species, the entanglement between the two sites and the entanglement between the two species. The latter is measured by means of the Schmidt gap, the von Neumann entropy or the purity obtained after tracing out a part of the system. A number of relevant states are identified, e.g. Schr\"odinger catlike many-body states, in which the outcome of the population imbalance of both components is completely correlated, and other states with even larger von Neumann entropy which have a large spread in Fock space. 
\end{abstract}

\pacs{
67.60.Bc 
67.85.Bc 
03.75.Gg 
}

\maketitle
\section{Introduction}

Bose-Einstein condensates trapped in double-well potentials are 
interesting not only from a fundamental point of 
view~\cite{Smerzi,Leggett,Gati} but also for their potential 
applications. Among the latter, the most prominent examples 
are found in quantum metrology~\cite{Esteve, Gross1, Riedel,zibold,Gross2}, 
and matterwave interferometry~\cite{Berrada1, Berrada2}. A crucial 
aspect of this system is that it can host relevant entangled 
many-body states, e.g. Schr\"odinger catlike 
states~\cite{cirac,BrunoArtur1,carr10,Mazzarella2,ma2}, or pseudo-spin 
squeezed states, as already demonstrated experimentally for 
the single component case~\cite{Esteve,Gross1,Riedel,Strobel,Muessel2,Muessel}. 

Going from the single component into the binary mixture case the richness 
of the possible many-body correlations is increased. For instance, to the 
spatial entanglement present in the single component trapped in the 
double-well we now add the possibility of having entanglement between 
the two species. The interplay between spatial and interspecies correlations 
allows one to have a variety of correlated many-body states depending on 
the atom-atom interactions and tunneling strengths. 

Our article explores the many-body properties of the binary mixture. 
Most of previous studies have concentrated on mean-field 
descriptions~\cite{ashhab,ng,wen,Xu,Satija,jul09,Mele2,Mazzarella} discussing 
dynamical features related to the Josephson to self-trapping transitions. 
Binary mixtures in the mean field approximation have also been studied 
in the context of measure synchronization~\cite{qiu0}. Beyond 
mean field studies include the onset of hybrid synchronization between 
a mean-field subsystem and a full quantum one~\cite{qiu2}, the dynamical generation 
of correlated states~\cite{Teichmann,Chatterjee,Kronke}, and also the 
extension of measure synchronization to many-body systems~\cite{qiu1}.

In this article we concentrate on fully characterizing the many-body 
properties of the binary mixture by means of quantum information tools, 
such as the entanglement spectrum and Schmidt gap~\cite{Chiara,Gallemi1, Gallemi2} 
of the ground state of the system. We will complement them by usual 
many-body techniques like the computation of the condensed fractions, 
population imbalances of the two species and ground state energy gap. 
A precise knowledge of the ground states which can appear for both 
attractive and repulsive atom-atom interactions may later be used to 
design protocols to produce desired many-body correlations, 
dynamically~\cite{Teichmann, Zibold, Muessel}, by control 
theory~\cite{Lapert,tichy} or by means of shortcut protocols~\cite{Torrontegui,Yuste,campbell}. 

The paper is organized as follows. First, in Sect.~\ref{sec2} we will describe the theoretical 
model and explain the procedure used to obtain the quantum many-body states. Then, in 
Sect.~\ref{sec3}, we  present the magnitudes used to characterize the ground state properties 
of the system. The main ones are: 1) the imbalance of population between the wells, 
which can be measured experimentally in single component bosons; 2) the condensed 
fraction, which measures the degree of Bose-Einstein condensation of each component, 
and 3) entanglement measures, like von Neumann entropies of the subsystems after 
bipartition. Sections \ref{sec4},~\ref{sec5} and \ref{sec6} contain the main results. 
In Sect.~\ref{sec4} we discuss 
the symmetric case, in which the bosons of the two species, $A$ and $B$, have the 
same intraspecies interaction. In Sect.~\ref{sec5} we consider a more general case, in which 
the intraspecies interactions are not taken equal. In Sect.~\ref{sec6} we explore the effect 
of having different number of particles of each species. Finally, in Sect.~\ref{sec7} we 
provide a brief summary and conclusions.
\section{Description of the model}
\label{sec2}

A mixture of two bosonic species with fixed number of particles, $N_A$ particles of $A$ and $N_B$ of $B$, is trapped in a double well potential. The atom-atom interaction is assumed to be well approximated by a contact potential. Further, we consider only two single particle modes for each species~\cite{Gati}.  Under this approximations we have the following second quantized Hamiltonian,

\begin{eqnarray}
\hat{H} &=& -J_A \left ({\hat{a}}^{\dagger}_R \hat{a}_L + {\hat{a}}^{\dagger}_L \hat{a}_R \right ) -J_B \left ({\hat{b}}^{\dagger}_R \hat{b}_L + {\hat{b}}^{\dagger}_L \hat{b}_R \right) \nonumber
\\
&+& \displaystyle \frac{U_{AA}}{2} \left [ \hat{n}^{A}_L (\hat{n}^{A}_L -1 ) + \hat{n}^{A}_R (\hat{n}^{A}_R - 1) \right ] \nonumber
\\
&+& \displaystyle \frac{U_{BB}}{2} \left [ \hat{n}^{B}_L (\hat{n}^{B}_L -1 ) +
\hat{n}^{B}_R (\hat{n}^{B}_R - 1) \right ] \nonumber
\\
&+& \displaystyle U_{AB} \left ( \hat{n}^{A}_L \hat{n}^{B}_L + \hat{n}^{A}_R \hat{n}^{B}_R \right )-\epsilon \left (\hat{n}^{B}_L - \hat{n}^{B}_R \right ) ,
\label{eq:1}
\end{eqnarray}
where $[ {\hat{a}}_i ,{\hat{a}}^{\dagger}_j ]=\delta_{i,j}$ , $[ {\hat{b}}_i ,{\hat{b}}^{\dagger}_j ]=\delta_{i,j}$, $\hat{n}^A_{i}=\hat{a}^{\dagger}_i\hat{a}_i$, $\hat{n}^B_{i}=\hat{b}^{\dagger}_i\hat{b}_i$ and $i,j=L,R$  ($L$ associated with the left site and $R$ with the right site). The action of the creation and annihilation operators on the Fock basis reads, for instance for L,
\begin{eqnarray}
{\hat{a}}^{\dagger}_L \ket{{n}^{A}_L,{n}^{A}_R,{n}^{B}_L,{n}^{B}_R} &=& \sqrt{{n}^{A}_L +1} \ket{{n}^{A}_L +1,{n}^{A}_R,{n}^{B}_L,{n}^{B}_R}, \nonumber
\\
{\hat{b}}^{\dagger}_L \ket{{n}^{A}_L,{n}^{A}_R,{n}^{B}_L,{n}^{B}_R} &=& \sqrt{{n}^{B}_L +1} \ket{{n}^{A}_L,{n}^{A}_R,{n}^{B}_L +1,{n}^{B}_R}, \nonumber
\\
{\hat{a}}_L \ket{{n}^{A}_L,{n}^{A}_R,{n}^{B}_L,{n}^{B}_R} &=& \sqrt{{n}^{A}_L} \ket{{n}^{A}_L -1,{n}^{A}_R,{n}^{B}_L,{n}^{B}_R}, \nonumber
\\
{\hat{b}}_L \ket{{n}^{A}_L,{n}^{A}_R,{n}^{B}_L,{n}^{B}_R} &=& \sqrt{{n}^{B}_L} \ket{{n}^{A}_L,{n}^{A}_R,{n}^{B}_L-1,{n}^{B}_R},
\label{eq:2}
\end{eqnarray}
where the Fock basis is characterized by the number of particles of each species in each one of the sites.
The strength of the intra, AA and BB, and interspecies, AB, interaction is given by the parameter $U_{AA}$, $U_{BB}$ and $U_{AB}$, respectively~\cite{Smerzi, Bloch}. Within our sign convention, positive and negative values of $U_{\alpha \beta}$ correspond to repulsive and attractive interactions, respectively. The hopping parameters $J_A$ and $J_B$ can in principle be varied by raising or lowering the potential barrier between the two wells. A small bias term, $0 < \epsilon \ll J_A$,$J_B$ , ensures the breaking of left$-$right symmetry and also $A-B$ symmetry. In our case it has been chosen to be energetically favourable to have $B$ particles on the L site. We define the parameters $\Lambda_A \equiv N_A U_{AA}/J_A$, $\Lambda_B \equiv N_B U_{BB}/J_B$ and $\Lambda_{AB} \equiv N_A U_{AB}/J_A=N_B U_{AB}/J_B$. \par
To diagonalize the Hamiltonian, the Fock basis used in (\ref{eq:2}) is labelled as
\begin{equation}
\label{eq:2.1}
\ket{k_A,k_B} \equiv \ket{N_A-k_A,k_A}\ket{N_B-k_B,k_B},
\end{equation}
where $k_A = 0,..., N_A$ and $k_B=0,...,N_B$ and thus the dimension of the Hilbert space is $(N_A+1)(N_B+1)$. The state $\ket{k_A,k_B}$ is the one having $k_A$ bosons of type $A$ on the right and $k_B$ bosons of type $B$ on the right.
\par Therefore, the matrix elements of the Hamiltonian  are,
\begin{eqnarray}
\label{eq:3}
&\bra{k'_A,k'_B}&\hat{H}\ket{k_A,k_B}= \nonumber \\ \nonumber
&&\frac{U_{AA}}{2}  [ (N_A-k_A)(N_A-k_A-1)\\\nonumber
&+&k_A(k_A-1) ] \delta_{k_A,k'_A}\delta_{k_B,k'_B}\\\nonumber
&+&\frac{U_{BB}}{2} [ (N_B-k_B)(N_B-k_B-1)\\\nonumber
&+&k_B(k_B-1) ] \delta_{k_A,k'_A}\delta_{k_B,k'_B}\\\nonumber
&-& J_A \sqrt{(N_A-k_A)(k_A+1)}\,\delta_{k_A+1,k'_A}\delta_{k_B,k'_B}\\\nonumber
&-&J_A \sqrt{k_A(N_A+1-k_A)}\,\delta_{k_A-1,k'_A}\delta_{k_B,k'_B}\\\nonumber
&-& J_B \sqrt{(N_B-k_B)(k_B+1)}\,\delta_{k_A,k'_A}\delta_{k_B+1,k'_B}\\\nonumber
&-&J_B \sqrt{k_B(N_B+1-k_B)}\,\delta_{k_A,k'_A}\delta_{k_B-1,k'_B}\\\nonumber
&+& U_{AB}  [(N_A-k_A)(N_B-k_B)\\\nonumber
&+&k_Ak_B]\delta_{k_A,k'_A}\delta_{k_B,k'_B}\\
&-&\epsilon(N_B -2k_B)\delta_{k_A,k'_A}\delta_{k_B,k'_B}.
\end{eqnarray}

Diagonalizing the Hamiltonian matrix introduced in (\ref{eq:3}), the energy spectrum is obtained numerically and the ground state is found in different situations. 
\section{Ground state properties}
\label{sec3}

\subsection{Spectral decomposition and degeneracy}
The ground state of the system $\ket{\Psi_{0}}$, can be expressed in the Fock basis as
\begin{equation}
\label{eq:4}
\ket{\Psi_{0}}=\sum_{k_{A}=0}^{N_A}\sum_{k_{B}=0}^{N_B}C_{k_A,k_B}\ket{k_A,k_B}.
\end{equation}
Since it is an eigenvector of the Hamiltonian it satisfies $\hat{H}\ket{\Psi_{0}}=E_0\ket{\Psi_{0}}$, with $E_0$ the energy of the ground state. The first excited state $\ket{\Psi_{1}}$ satisfies $\hat{H}\ket{\Psi_{1}}=E_1\ket{\Psi_{1}}$. Degeneracy will occur when at least two different eigenstates have the same energy. For this reason, the difference 
\begin{equation}
\label{eq:4.2}
\Delta E_{1,0}\equiv E_1 - E_0 ,
\end{equation}
determines whether the ground state is degenerate or not.
\subsection{Population imbalance}
The population imbalance $z_i$, with $i=A,B$, for a given arbitrary state of the system $\ket{\Psi}$, is defined as the expectation value
\begin{equation}
\label{eq:5}
z_i\equiv \frac{1}{N_i}\bra{\Psi}{\hat{n}_L}^{i}-{\hat{n}_R}^{i}\ket{\Psi}.
\end{equation}
For a state with all particles of type $A$($B$) on the left site $z_{A(B)}$ is $1$, while if all the particles are on the right site its value is $-1$. This quantity is zero in the case of equal population of particles of a given type in the two sites. We can also compute the dispersion of $z_i$
\begin{equation}
\label{eq:6}
\sigma_z^i\equiv \displaystyle \sqrt{
\bra{\Psi}
\left(\frac{{\hat{n}_L}^{i}-{\hat{n}_R}^{i}}{N_i}\right)^2 \ket{\Psi}
- \left (\bra{\Psi}
\frac{{\hat{n}_L}^{i}-{\hat{n}_R}^{i}}{N_i}\ket{\Psi} \right )^2
}.
\end{equation}
Particularly for the ground state $\ket{\Psi_0}$ using its spectral decomposition (\ref{eq:4}) we can express the population imbalance (\ref{eq:5}) for each component of the mixture as

\begin{equation}
\label{eq:7}
z_i=\sum_{k_{A}=0}^{N_A}\sum_{k_{B}=0}^{N_B}|C_{k_A,k_B}|^2\frac{N_i-2k_i}{N_i}
\end{equation}
and also the corresponding dispersion (\ref{eq:6}) as

\begin{eqnarray}
\label{eq:8}
\sigma_z^i&=&\left[
\sum_{k_{A}=0}^{N_A}\sum_{k_{B}=0}^{N_B}|C_{k_A,k_B}|^2 
\left(\frac{N_i-2k_i}{N_i}\right)^2 \nonumber \right.\\
&-&  \left. \left(\sum_{k_{A}=0}^{N_A}\sum_{k_{B}=0}^{N_B}|C_{k_A,k_B}|^2
\frac{N_i-2k_i}{N_i}\right)^2 
\right]^{1/2} .
\end{eqnarray}

\subsection{Degree of condensation}
The degree of condensation of each of the species in the ground state is characterized using the one-body density matrix for each component

\begin{equation}
\label{eq:9}
\rho_{i,j}^A\equiv \frac{1}{N_A}\bra{\Psi_0}\hat{a}^{\dagger}_i\hat{a}_j\ket{\Psi_0},
\end{equation}
\begin{equation}
\label{eq:10}
\rho_{i,j}^B\equiv \frac{1}{N_B}\bra{\Psi_0}\hat{b}^{\dagger}_i\hat{b}_j\ket{\Psi_0} ,
\end{equation}
where in both equations (\ref{eq:9}) and (\ref{eq:10}) $i,j=L,R$. \par Diagonalizing $\rho^{A(B)}$ we obtain its eigenvalues $n^{A(B)}_1$ and $n^{A(B)}_2$ normalized to unity, which correspond to the occupations of the single-particle eigenstates of the one-body density matrix $\ket{\phi_1^{A(B)}}$ and $\ket{\phi_2^{A(B)}}$ respectively.
\par The eigenvalues fulfill $n^{A(B)}_1+n^{A(B)}_2=1$ and, by definition, $0\le n^{A(B)}_2 \le n^{A(B)}_1$. In the particular case when $n^{A(B)}_1=1$, all the bosons of type $A$($B$) populate the same single particle state $\ket{\phi_1^{A(B)}}$ and the state of the subsystem of bosons of type $A$($B$) can be written as a product state $\ket{\Phi_1^{A(B)}}\equiv\ket{\phi_1^{A(B)}}^{\otimes N_{A(B)}}$. 
\subsection{Partial traces, purity and entanglement}
The density matrix associated with the ground state $\ket{\Psi_0}$, that describes completely the state of the total system formed by the two types of particles, is

\begin{equation}
\label{eq:11}
\hat{\rho}_0\equiv\ket{\Psi_0}\bra{\Psi_0} .
\end{equation}
This matrix has dimension $(N_A+1)(N_B+1)$. If we are interested in only one part of the system, for instance the type $A$($B$) bosons, we can obtain the state for this subsystem taking the partial trace with respect to $B$($A$) of the matrix $\hat{\rho}_0$. The state of type $A$($B$) bosons then would be described by $\hat{\rho}^{A(B)}_0$, that is

\begin{equation}
\label{eq:12}
\hat{\rho}^{A(B)}_0\equiv {\rm Tr}_{B(A)}[\hat{\rho}_0] ,
\end{equation}
which has dimension $(N_{A(B)}+1)$.
\par In general, after tracing out part of the system, the state of the remaining subsystem is a mixed state. In order to determine if $\hat{\rho}^{A(B)}_0$ is a pure state or not and its degree of purity, the trace of this matrix squared, $P_{A(B)}$, is computed,

\begin{equation}
\label{eq:13}
P_{A(B)}\equiv {\rm Tr}[(\hat{\rho}^{A(B)}_0)^{2}].
\end{equation}
\par
When the state of each subsystem is a pure state we obtain $P_{A(B)}=1$. In this case the ground state is a product state $\ket{\Psi_0}=\ket{\Psi_0^A}\ket{\Psi_0^B}$ and there is no entanglement between $A$ and B. Otherwise, when $P_{A(B)}\ne1$, the ground state is not a product state,$\ket{\Psi_0}\ne\ket{\Psi_0^A}\ket{\Psi_0^B}$, and $P_{A(B)}$ satisfies $\frac{1}{N_{A(B)}+1} \le P_{A(B)}<1$, now having entanglement between $A$ and $B$ with completely entangled subsystems for the case $\hat{\rho}^{A(B)}_0=\frac{1}{N_{A(B)}+1}\mathbb{I}$.
\subsection{Entropy and Schmidt gap}
The von Neumann entropy of the state of each subsystem is computed diagonalizing

\begin{equation}
\label{eq:14}
\hat{\rho}^{A(B)}_0=\sum_{i=0}^{N_{A(B)}}\lambda^{A(B)}_i\ket{\lambda^{A(B)}_i}\bra{\lambda^{A(B)}_i},
\end{equation}
to obtain its eigenvalues $\lambda^{A(B)}_i$, considering $\lambda^{A(B)}_0 \ge \lambda^{A(B)}_1 \ge ... \ge \lambda^{A(B)}_{N_{A(B)}}$. As the density matrix $\hat{\rho}^{A(B)}_0$ is normalized, ${\rm Tr}[\hat{\rho}^{A(B)}_0]=1$, the eigenvalues satisfy $\sum_{i=0}^{{N_{A(B)}}} \lambda^{A(B)}_i = 1$. In order to calculate the entropy $S_{A(B)}$ we use the definition

\begin{equation}
\label{eq:15}
S_{A(B)}\equiv-{\rm Tr}\left[\hat{\rho}^{A(B)}_0 \log \hat{\rho}^{A(B)}_0\right]
=-\sum_{i=0}^{N_{A(B)}}\lambda^{A(B)}_i \log\lambda^{A(B)}_i ,
\end{equation}
where if a $\lambda^{A(B)}_i = 0$, the corresponding term is considered to be zero and it is not added to the sum. The entropy has a minimum value equal to zero when all the $\lambda^{A(B)}_i = 0$ except $\lambda^{A(B)}_0 = 1$ and then the state is pure. Its maximum value, $S_{max}^{A(B)}= {\log (N_{A(B)}+1)}$, is reached when $\lambda_0 = \lambda^{A(B)}_1 = ... = \lambda^{A(B)}_{N_{A(B)}} = \frac{1}{N_{A(B)}+1}$ and in this case we have the maximum entanglement situation (discussed previously) which means that each subsystem is in a mixed state. In fact, as the density matrix of the system (\ref{eq:11}) always corresponds to a pure state in our case, $S_A=S_B$. This is derived from the triangle inequality that relates the von Neumann entropy of the whole system and the partial entropies \cite{Araki}. For this reason, if $N_A\neq N_B$ then  $S_{max}^{A(B)}= \min \{{\log (N_{A}+1)}, {\log (N_{B}+1)}\}$.
\par
The Schmidt gap is defined as the difference between the two largest eigenvalues of the density matrix,

\begin{equation}
\Delta \lambda^{A(B)} \equiv \lambda^{A(B)}_0 - \lambda^{A(B)}_1 ,
\end{equation}
and also distinguishes between having pure states and no entanglement between the $A$ and $B$ when $\Delta \lambda^{A(B)} = 1$ and totally entangled subsystems and mixed states for part $A$ and $B$ if  $\Delta \lambda^{A(B)} = 0$.
\par
Moreover, $L-R$ quantum correlations are quantified by tracing out the L(R) part of the system,
\begin{equation}
\label{eq:19}
\hat{\rho}^{L(R)}_0\equiv {\rm Tr}_{R(L)}[\hat{\rho}_0] ,
\end{equation}
and computing the von Neumann entropy defined as
\begin{equation}
S_{LR}\equiv-{\rm Tr}\left[\hat{\rho}^{L}_0 \log \hat{\rho}^{L}_0\right]
= -{\rm Tr}\left[\hat{\rho}^{R}_0 \log \hat{\rho}^{R}_0\right] ,
\end{equation}
that using the Fock basis [\ref{eq:2.1}] reads
\begin{equation}
\label{eq:21}
S_{LR}=-\sum_{k_A=0}^{N_A}\sum_{k_B=0}^{N_B}|C_{k_A,k_B}|^2 \log |C_{k_A,k_B}|^2 .
\end{equation}
\par
We also characterize $L-R$ quantum correlations within each species tracing out the L(R) part in each subsystem (\ref{eq:14}) and computing the von Neumann entropy for A as
\begin{equation}
S_{LR}^{A}=-\sum_{k_A=0}^{N_A} \left[\left ( \sum_{k_B=0}^{N_B}|C_{k_A,k_B}|^2 \right ) \log \left ( \sum_{k_B=0}^{N_B}|C_{k_A,k_B}|^2 \right ) \right],
\end{equation}
and for species $B$ as
\begin{equation}
S_{LR}^{B}=-\sum_{k_B=0}^{N_B} \left[\left ( \sum_{k_A=0}^{N_A}|C_{k_A,k_B}|^2 \right ) \log \left ( \sum_{k_A=0}^{N_A}|C_{k_A,k_B}|^2 \right ) \right].
\end{equation}
The maximum values of $L-R$ entropies are $S^{LR}_{max} = \log[(N_A+1)(N_B+1)]$ and $S^{A(B)}_{LR_{max}}= \log(N_{A(B)}+1)$.
\section{Equal intraspecies interaction}
\label{sec4}

In this section we present our results for the case in which the intraspecies interaction is the same for both species. We discuss how the properties of the system change as we vary the interspecies one, $U_{AB}$. In terms of our parameters, we concentrate in the cases $N_A=N_B$, $U_{AA}=U_{BB}$, $J_A=J_B$ and $U_{AB}\ne U_{AA}$. Now for simplicity we define $\Lambda\equiv\Lambda_A=\Lambda_B$ and also $J \equiv J_A=J_B$. In our numerical calculations $N_A=N_B=20$, $J=20$ and $\epsilon/J=10^{-10}$. Thus, the Hamiltonian is a $441\times441$ matrix, which is diagonalized for different values of $\Lambda$ and $\Lambda_{AB}$. We organize the section as follows. First we discuss the spectral decomposition of the ground state depending on the character of the intraspecies interaction. Second, we study the condensed fraction and entanglement properties as a function of $\Lambda$ and $\Lambda_{AB}$.
\subsection{Repulsive intraspecies interaction}
In this case we have $\Lambda>0$, i.e. atoms of each species repel each other. We can consider three situations, $\Lambda_{AB}>0$, $\Lambda_{AB}<0$ and $\Lambda_{AB}=0$.

\begin{table*}[t]
\centering
\begin{tabular}{ |c|c|c| }
\hline
 & $\Lambda>0$ & $\Lambda<0$ \\
 \hline
 $\Lambda_{AB}>0$ & $(1/ \sqrt{2})(\ket{N_A,0}\pm \ket{0,N_B})$ & $(1/\sqrt{2})( \ket{N_A,0}\pm \ket{0,N_B}) $  \\
 \hline
 $\Lambda_{AB}=0$ $(|\Lambda| \gg 0)$ & $\ket{N_A/2,N_B/2}$   & $(1/2)(\ket{N_A,N_B}+ \ket{N_A,0}+\ket{0,N_B}+\ket{0,0})$ \\
 \hline
 $\Lambda_{AB}<0$ & $(1/ \sqrt{2})(\ket{0,0} \pm \ket{N_A,N_B}) $ &  $(1/ \sqrt{2})( \ket{0,0}\pm \ket{N_A,N_B}) $   \\ 
 \hline
\end{tabular}
\caption{Ground state for the case $|\Lambda_{AB}| \gg |\Lambda|$ in absence of tunnelling ($|\Lambda| \gg 1$) and in absence of bias depending on the character of inter and intraspecies interaction.}
\end{table*}

\begin{figure}[t]
\centering
\resizebox{\columnwidth}{!}{\input{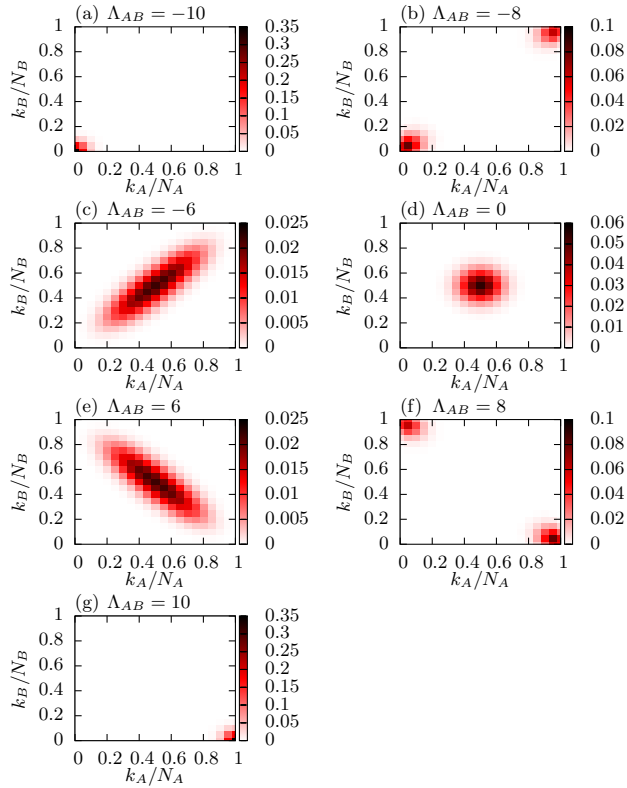}}
\caption{Spectral decomposition of the ground state, $|C_{k_A,k_B}|^2$, for repulsive intraspecies interaction ($\Lambda>0$), plotted for different values of $\Lambda_{AB}$. We observe the transition from (a), a regime dominated by the attraction between the different species to a regime (g) dominated by the repulsion between species $A$ and $B$, going through intermediate regimes in the panels (b) to (f). For all panels $\Lambda=4$, $N_A=N_B=20$, $J=20$ and $\epsilon/J=10^{-10}$.}
\label{fig:1}
\end{figure}

${\bf \Lambda_{AB}>0}$. In this case, particles of different type do not want to be at the same site. For $\Lambda_{AB} \gg \Lambda \gg 1$, that is neglecting tunnelling effects, this can be achieved with all type $A$ bosons on the right site and type $B$ bosons on the left site, or similarly with type $B$ bosons on the right and type $A$ bosons on the left site. These two states are degenerate in this limit and, therefore, any superposition of them, $\ket{\Psi_0} =\alpha \ket{N_A,0} + \beta \ket{0,N_B} $, with $|\alpha|^2+|\beta|^2=1$, has the same energy. This degeneracy is broken by the bias term introduced in the Hamiltonian, and since the bias has been chosen to be energetically favourable to have $B$ particles on the left site, the ground state is $\ket{\Psi_0} = \ket{N_A,0}$. This limiting case, obtained for $\Lambda_{AB} \gg \Lambda \gg 1$, can help to understand the numerical results. For instance in Fig.~\ref{fig:1}(g) we present the spectral decomposition of the state obtained for $\Lambda_{AB}=10$ and $\Lambda=4$. The ground state is seen to be close to the $\ket{N_A,0}$. As the $\Lambda_{AB}$ is decreased, the bias term is not large enough to localize the state and the ground state is closer to the linear combination $\frac{1}{\sqrt{2}}(\ket{N_A,0} \pm \ket{0,N_B})$, 
see Fig.~\ref{fig:1}(f).

${\bf \Lambda_{AB}<0}$. In absence of tunnelling, attractive interaction between different bosons, will make all of them be at the same site despite the repulsion between same type bosons because now we consider the regime where $-\Lambda_{AB}\gg \Lambda \gg 1$. There is degeneracy in this case, too. Bosons can be all together on the right, $\ket{\Psi_0} = \ket{N_A,N_B}$, or on the left $\ket{\Psi_0} = \ket{0,0}$, or on any superposition $\ket{\Psi_0} = \alpha \ket{N_A,N_B} + \beta \ket{0,0} $, with $|\alpha|^2+|\beta|^2=1$. Again, the bias breaks the symmetry and in this case selects the state $\ket{\Psi_0} = \ket{0,0}$ because type $B$ bosons are on the left site in this one. The numerical results do agree with these arguments. For $\Lambda_{AB}=-10$ the ground state is close to the state $\ket{0,0}$, see 
Fig.~\ref{fig:1}(a). For $\Lambda_{AB}=-8$ the bias does not localize the state and the ground state is close to a linear combination $\frac{1}{\sqrt{2}}(\ket{0,0} \pm \ket{N_A,N_B})$.
\par

For ${\bf \Lambda_{AB}=0}$, the system is equivalent to having two independent bosonic Josephson junctions, which only differ by the presence of a bias in one of them (in the $B$ component). Single-component Bose-Einstein condensates in a double-well have been studied in detail in~\cite{Smerzi,cirac, Gati, BrunoArtur1,ma2}. The ground state of the full system is the direct product of the ground state of each subsystem. Thus, we obtain a left-right $A-B$ symmetric ground state (see Fig.~\ref{fig:1}(d)). This type of state would be binomial for each component~\cite{BrunoArtur1} for $\Lambda=0$. For $\Lambda>0$ it becomes squeezed. In the limit case when $|\Lambda|\gg1$  it tends to $\ket{\Psi_0} = \ket{\frac{N_A}{2},\frac{N_B}{2}}$.

When the bias does not play a role, i.e. when $\epsilon \ll \Delta E_{1,0}$, the Hamiltonian has left-right and $A-B$ symmetries so in this situation its eigenstates have these symmetries too. Therefore, catlike states appear for $\Lambda_{AB}<0$ as well as for $\Lambda_{AB}>0$. In both cases, the ground state is quasi-degenerate with the first excited state. Without bias in the $\Lambda_{AB}$ dominated regime we have for attractive intraspecies interaction $\ket{\Psi_0} = \frac{1}{\sqrt{2}} \left ( \ket{N_A,N_B} \pm \ket{0,0} \right )$ and for the repulsive case $\ket{\Psi_0} = \frac{1}{\sqrt{2}} \left ( \ket{N_A,0} \pm \ket{0,N_B} \right )$. This is mainly what is found in Fig.~\ref{fig:1}(b) and Fig.~\ref{fig:1}(f) respectively. For $|\Lambda_{AB}|$ small enough, the two peaks merge and form a broadened peak which has its tails pointing to the previous corresponding two peaks (see Figs.~\ref{fig:1}(c) and ~\ref{fig:1}(e)) and becomes narrow when $|\Lambda_{AB}|$ decreases until $\Lambda_{AB}=0$. As we will discus in subsection C, the states with larger spread, such as Figs.~\ref{fig:1}(c) and~\ref{fig:1}(e) will have the larger entropy, marking regions where the ground state goes from localized to highly delocalized catlike states (see Figs.~\ref{fig:1}(b) and~\ref{fig:1}(f)).
\subsection{Attractive intraspecies interaction}
Again, we can distinguish three cases: $\Lambda_{AB}>0$, $\Lambda_{AB} <0$ and $\Lambda_{AB}=0$.

\begin{figure}[t]
\centering
\resizebox{\columnwidth}{!}{\input{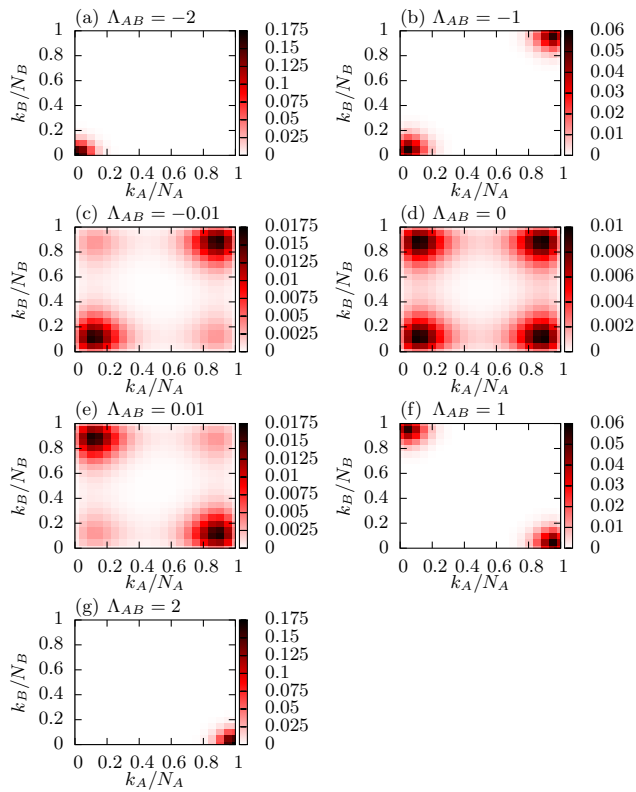}}
\caption{Spectral decomposition of the ground state, $|C_{k_A,k_B}|^2$, for attractive intraspecies interaction ($\Lambda<0$), plotted for different values of $\Lambda_{AB}$. We observe the transition from (a), a regime dominated by the attraction between the different species to a regime (g) dominated by the repulsion between species $A$ and $B$, going through intermediate regimes in the panels (b) to (f). For all panels $\Lambda=-3$, $N_A=N_B=20$, $J=20$ and $\epsilon/J=10^{-10}$.}
\label{fig:2}
\end{figure}

${\bf \Lambda_{AB}>0}$. Here the repulsion between different type bosons and the attraction between same type bosons are not competing, in the sense that both effects can be easily fulfilled simultaneously, which did not happen in the two first limits discussed in subsection A. Particles of the same type want to be together and separated from the other type ones. The states that accomplish this in absence of tunnelling are  $\ket{\Psi_0} = \ket{N_A,0}$, $\ket{\Psi_0} = \ket{0,N_B}$ and their superposition $\ket{\Psi_0} =\alpha \ket{N_A,0} + \beta \ket{0,N_B}$ , with $|\alpha|^2+|\beta|^2=1$. The bias breaks the symmetry towards $\ket{\Psi_0} = \ket{N_A,0}$ (see Fig.~\ref{fig:2}(g)). Notice that this argument also holds for the repulsive-repulsive case (repulsion between same and different type of bosons).

${\bf \Lambda_{AB}<0}$. This is the attractive-attractive case where, in absence of tunnelling, it is clear that all the bosons will be at the same site. Now the states expected are the same as in the regime $-\Lambda_{AB} \gg \Lambda \gg 1$. The difference in this case is that the effects of $\Lambda$ and $\Lambda_{AB}$ go in the same direction. The ground state candidates are $\ket{\Psi_0} = \ket{N_A,N_B}$, $\ket{\Psi_0} = \ket{0,0}$ and, as before, their superposition $\ket{\Psi_0} = \alpha \ket{N_A,N_B} + \beta \ket{0,0}$, with $|\alpha|^2+|\beta|^2=1$, with the bias breaking the symmetry and selecting the state $\ket{\Psi_0} = \ket{0,0}$ 
(see Fig.~\ref{fig:2}(a)).

Here, we also have the catlike states described in the previous section as shown in Figs.~\ref{fig:2}(b) and~\ref{fig:2}(f). However, the ground state for ${\bf \Lambda_{AB}=0}$ is a different one (see Fig.~\ref{fig:2}(d)) and thus the intermediate states too (see Fig.~\ref{fig:2}(c) and~\ref{fig:2}(e)). The ground state for zero intraspecies interaction is degenerate not only with the first excited state but also with the second, the third and the fourth. Looking only in one type of bosons we have in these conditions a left-right catlike state 
\cite{cirac,BrunoArtur1}, so here we see this for both components at the same time. Each catlike state for each species is degenerate, what means that the ground state of the whole system is a linear combination of $\ket{0,0}$, $\ket{N_A,0}$, $\ket{0,N_B}$ and $\ket{N_A,N_B}$. As occurred before, the tunnelling mixes the four states and for finite $\Lambda$ we obtain states close to these ones but with a finite width in the Fock space, Fig.~\ref{fig:2}(c,d,e). 

\begin{figure}[t]
\centering
\resizebox{\columnwidth}{!}{\input{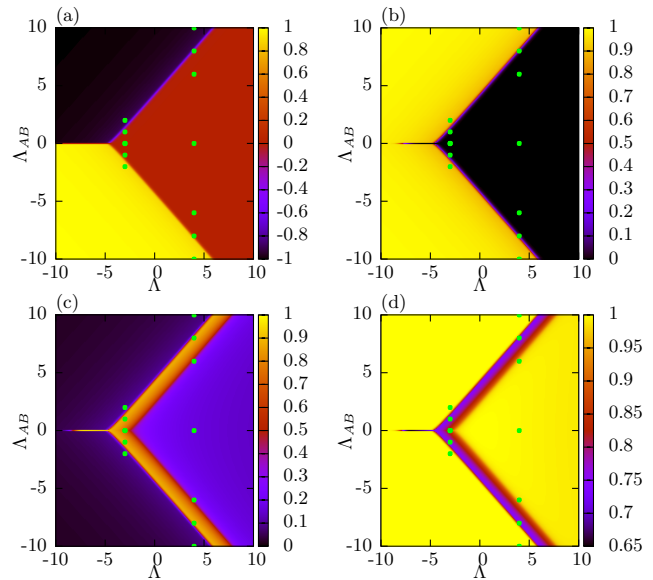}}
\caption{Properties of the ground state varying $\Lambda$ and $\Lambda_{AB}$. (a) Population imbalance, $z_A$. The particles of this type are all on the left $z_A=1$ (yellow), all on the right $z_A=-1$ (black) and $z_A=0$ (red). (b) Population imbalance, $z_B$. The particles of this type are all on the left $z_B=1$ (yellow) and $z_B=0$ (black). (c) Dispersion of population imbalance, $\sigma z_B$. (d) Condensed fraction, $n^B_1$. For $\Lambda_{AB}=0$ the results obtained in \cite{BrunoArtur1} are reproduced. Green spots mark where the states of Fig.~\ref{fig:1} and Fig.~\ref{fig:2} are found, except for the cases $\Lambda_{AB}=-0.01$ and $0.01$ which are not marked. For all panels $N_A=N_B=20$, $J=20$ and $\epsilon/J=10^{-10}$.}
\label{fig:3}
\end{figure}

In Table I we summarize the ground state in the interaction dominated regime.

\begin{figure}[t!]
\centering
\resizebox{\columnwidth}{!}{\input{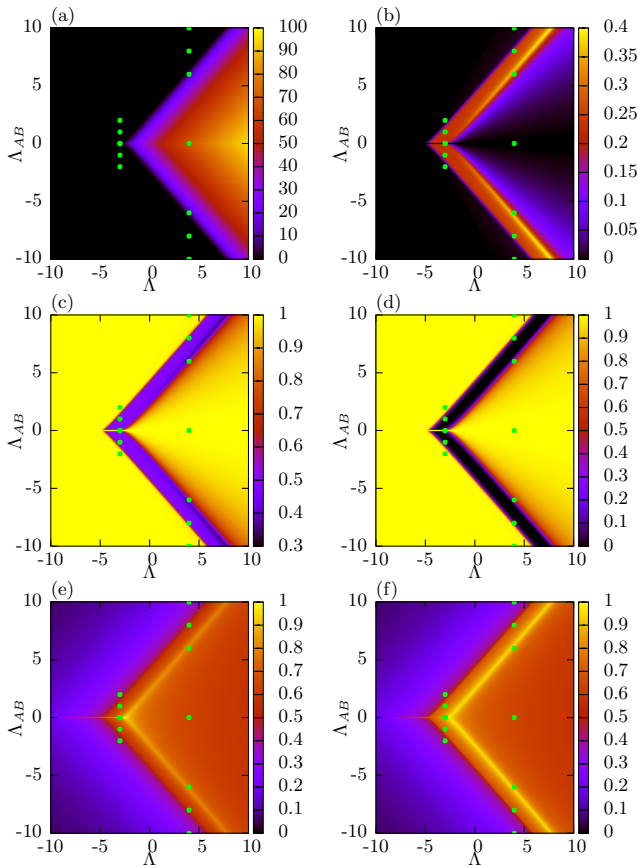}}
\caption{Properties of the ground state varying $\Lambda$ and $\Lambda_{AB}$. (a) Energy gap, $\Delta E_{1,0}$. The black region is where there is degeneracy. (b) Von Neumann entropy for subsystem $A$($B$), $S_{A(B)}$, normalized to its maximum value. (c) Trace of the density matrix squared for subsystem $A$($B$), $P_{A(B)}$. (d) Schmidt gap for subsystem $A$($B$), $\Delta \lambda^{A(B)}$. (e) L$-$R von Neumann entropy for the whole system, $S_{LR}$, and (f) for subsystem $A$($B$), $S_{LR}^{A(B)}$, both normalized to their maximum values. Green spots mark where the states of Fig.~\ref{fig:1} and Fig.~\ref{fig:2} are found, except for the cases $\Lambda_{AB}=-0.01$ and $0.01$ which are not marked. For all panels $N_A=N_B=20$, $J=20$ and $\epsilon/J=10^{-10}$.}
\label{fig:4}
\end{figure}

\subsection{Condensed fractions and entanglement properties}

Up to now, we have mainly discussed the spectral structure of the ground state 
for different values of $\Lambda$ and $\Lambda_{AB}$. Starting from the interaction 
dominated cases, we have understood the Fock space structure obtained in our numerical diagonalization. In particular, we have identified regimes
in which ground state quasidegeneracies and many-body fragmentation is expected to appear. In this section, we will characterize the condensation of the two ultracold atomic clouds and the entanglement between the two species.

First, let us provide a global picture and consider the population imbalance of the ground state for the $A$ and $B$ species as a function of $\Lambda$ and $\Lambda_{AB}$. The main difference between the two species is found in the attractive intraspecies interaction case, as seen by comparing panels (a) and (b) of Fig.~\ref{fig:3}. This is due to the bias term which breaks explicitly the $A-B$ symmetry, which in the interaction dominated regime localizes the $B$ atoms on the left and the $A$ atoms on the right. 

The population imbalance only provides an average information, which does not allow to differentiate for instance two very different quantum states, e.g. Fig.~\ref{fig:1}(b) and Fig.~\ref{fig:1}(d). Both of these states have a zero population imbalance, but the structure in Fock space is completely different. For instance, it can be inferred directly from the figure that the two states should have a very different quantum uncertainty for the imbalance of population. This means for instance the following: If one prepares the system in the state shown in Fig.~\ref{fig:1}(b) and measures the populations, the outcome of the measurements will be very polarized, i.e. almost all particles of each species will be found on the same well in each experiment. On average, however, we should find an average of population equal to zero. In contrast, in the state depicted in Fig.~\ref{fig:1}(d), the outcome of each individual experiment will almost never be too polarized, finding outcomes where a similar number of particles of each species is found in each well. These two states can be discriminated by means of the dispersion of the population imbalance $\sigma_z$, which is depicted in Fig.~\ref{fig:3}(c). 

For species $B$, the sharp lines delimiting the different population imbalance regions seen in Fig.~\ref{fig:3}(b) are replaced by broad transition regions in Fig.~\ref{fig:3}(c). This is a quantal effect, similar to the transition observed in the single component case \cite{zibold,BrunoArtur1}. As occurred in the single component case, in the transition regions the many-body state is very fragmented. As seen in Fig.~\ref{fig:3}(d), the $B$ component is almost fully condensed for all values of $\Lambda$ and $\Lambda_{AB}$ except for the transition regions, where the condensed fraction falls below 0.7. These fragmented states 
are for instance the ones in Fig.~\ref{fig:1}(b,c,e,f).

As explained above, in certain limits the ground state becomes degenerate with the first excited state. In Fig.~\ref{fig:4} (a) we depict $\Delta E_{1,0}$. Small degeneracies are not seen in the figure, and thus, for instance, the localisation due to the bias is not reflected in the figure. For $\Lambda<0$  the ground state is mostly degenerate. Gapped ground states are found for repulsive intraspecies interactions and also in the transition regions. 

The entanglement between the two species is characterized by the purity, the von Neumann entropy and the Schmidt gap of the density matrix after tracing out one of the species. The Schmidt gap provides a broad picture of presence of entanglement, see Fig.~\ref{fig:4}(d). As it only involves the difference between the two largest Schmidt coefficients, it does not differentiate between different entangled states, for instance, it cannot discriminate between a catlike state, Fig.~\ref{fig:1}(b) and a broadened peak Fig.~\ref{fig:1}(c). These two states can be told apart by computing the von Neumann entropy, Fig.~\ref{fig:4}(b). In this case, the broadened peak has a larger number of sizeable Schmidt coefficients than in the catlike case, and thus shows as a maximum of the von Neumann entropy, in yellow inside the transition region. A similar discussion can be made with the states in Fig.~\ref{fig:1}(e) and (f), which again have a similar Schmidt gap but different von Neumann entropy. The purity, shown in Fig.~\ref{fig:4}(c) provides a very similar global picture as the Schmidt gap.

The presence of quantum correlations between the two sites is shown in Fig.~\ref{fig:4}(e) and in Fig.~\ref{fig:4}(f) where $L-R$ von Neumann entropies are depicted, for the whole system and for just one single species, respectively. These two panels look similar but, for the whole system (Fig.~\ref{fig:4}(e)), the maximum value achieved of the entropy fixing $\Lambda_{AB}$ depends on the value of $\Lambda_{AB}$, whereas this dependence is not seen in $S_{LR}^{A(B)}$ shown in Fig.~\ref{fig:4}(f). Notice that in this particular case $S_{LR}^{A}=S_{LR}^{B}$. Due to the tunneling, we have quantum correlations between the two sites in absence of interaction between bosons of different type ($\Lambda_{AB}=0$) and also in the non-interacting case ($\Lambda=0$ and $\Lambda_{AB}=0$).

\section{Different intraspecies interaction}
\label{sec5}

In this section we will study the case $U_{AA}\ne U_{BB}$ with the same number of particles for each species, $N_A=N_B$ and as before $J\equiv J_A=J_B$ with $\epsilon/J=10^{-10}$. We consider a fixed value of $\Lambda_A$ for repulsive ($\Lambda_A>0$) and attractive interaction ($\Lambda_A<0$) and allow for variations of the parameters $\Lambda_B$ and $\Lambda_{AB}$.
\subsection{Repulsive intraspecies interaction $\Lambda_A$}
\begin{figure}[t]
\centering
\resizebox{\columnwidth}{!}{\input{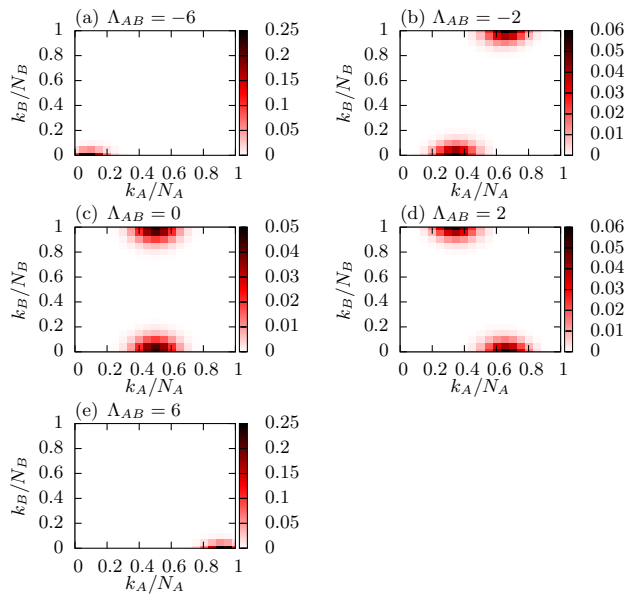}}
\caption{Spectral decomposition of the ground state, $|C_{k_A,k_B}|^2$, plotted for different values of $\Lambda_{AB}$. Here there is repulsion between $A$ ($\Lambda_A>0$) and attraction between $B$ type bosons ($\Lambda_B<0$). We observe the transition from (a), a regime dominated by the attraction between the different species to a regime (e) dominated by the repulsion between species $A$ and $B$, going through intermediate regimes in the panels (b) to (d). For all panels $\Lambda_A=4$, $\Lambda_B=-5$, $N_A=N_B=20$, $J=20$ and $\epsilon/J=10^{-10}$.}
\label{fig:5}
\end{figure}

\begin{figure}[t]
\centering
\resizebox{\columnwidth}{!}{\input{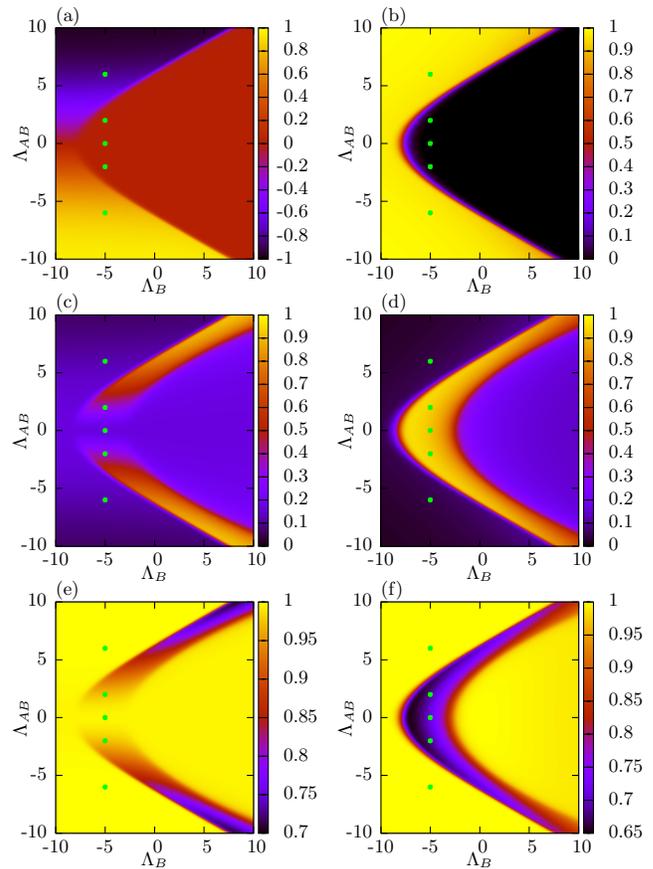}}
\caption{Properties of the ground state for $\Lambda_A>0$ and varying $\Lambda_B$ and $\Lambda_{AB}$. (a) Population imbalance $z_A$ and (b) $z_B$. Dispersion of population imbalance for each species, (c) $\sigma z_A$ and (d) $\sigma z_B$. Condensed fractions (e) $n^A_1$ and (f)  $n^B_1$. Green spots mark where the states 
of Fig.~\ref{fig:5} are found. For all panels $\Lambda_A=4$, $N_A=N_B=20$, $J=20$ and $\epsilon/J=10^{-10}$.}
\label{fig:6}
\end{figure}

\par
\begin{figure}[t]
\centering
\resizebox{\columnwidth}{!}{\input{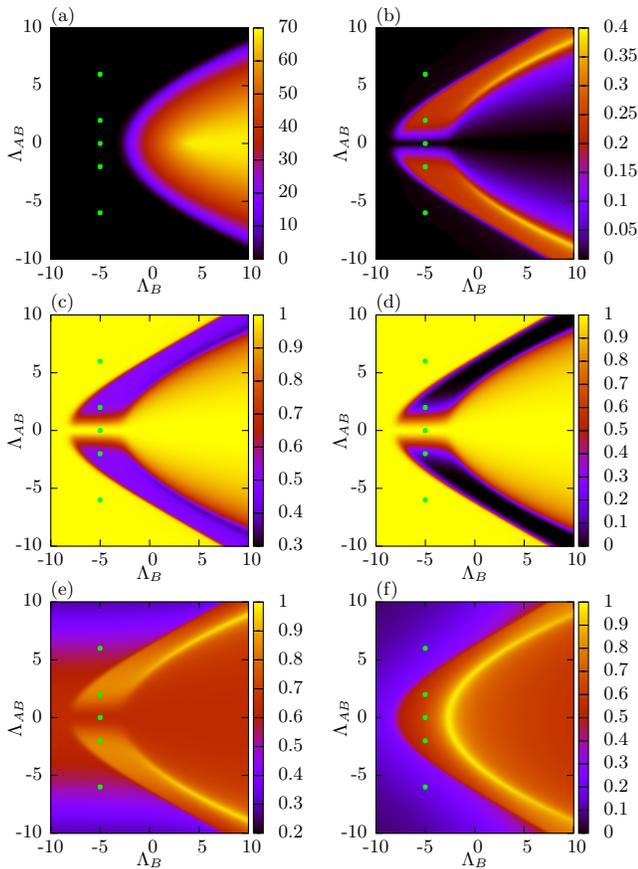}}
\caption{Properties of the ground state for $\Lambda_A>0$ and varying $\Lambda_B$ and $\Lambda_{AB}$. (a) Energy gap,  $\Delta E_{1,0}$. (b) Von Neumann entropy for subsystem $A$($B$), $S_{A(B)}$ normalized to its maximum value. (c) Trace of the density matrix squared for subsystem $A$($B$), $P_{A(B)}$. (d) Schmidt gap for subsystem $A$($B$), $\Delta \lambda^{A(B)}$. (e) L$-$R von Neumann entropy for subsystem $A$, $S_{LR}^A$, and (f) for subsystem $B$, $S_{LR}^{B}$, both normalized to their maximum values. Green spots mark where the states of Fig.~\ref{fig:5} are found. For all panels $\Lambda_A=4$, $N_A=N_B=20$, $J=20$ and $\epsilon/J=10^{-10}$.}
\label{fig:7}
\end{figure}

 Here we have found mainly the same type of states and transitions described in Sect.~\ref{sec4}. There are, however, new states and different behaviour for each type of bosons. For instance, for $\Lambda_B>0$ the states that are found are the ones of Fig.~\ref{fig:1} but appearing for different values of $\Lambda_{AB}$. For this reason we focus on the case of having $\Lambda_{B}<0$ as it is shown in Fig.~\ref{fig:5}. In the extreme cases (see Figs.~\ref{fig:5}(a) and~\ref{fig:5}(e)), the states are similar to the ones in Figs.~\ref{fig:1}(a) and ~\ref{fig:1}(g). In these cases the physics is dominated by $\Lambda_{AB}$ and the effect of the bias. Notice that, the different intraspecies interaction plays a relevant role because without the interspecies interaction, we would have a binomial-like distribution for type $A$ bosons (slightly repulsive intraspecies interaction, $\Lambda_A$) and a catlike state for type $B$ due to an attractive interaction in species $B$ (see Fig.~\ref{fig:5}(c)). When $\Lambda_{AB}$ increases (see Fig.~\ref{fig:5}(d)) or decreases (see Fig. ~\ref{fig:5}(b)) produces an entanglement of the A coefficients maintaining a catlike structure for the $B$ component, until a catlike state between the components $A$ and $B$ is formed. The states of Fig.~\ref{fig:5} are present along a vertical line in Figs.~\ref{fig:6} and ~\ref{fig:7} for a fixed value of $\Lambda_B$.

\par Existing differences between the two components are reflected in Fig.~\ref{fig:6} where we compare several observables. For the population imbalance of each species, panels (a) and (b), we obtain a similar description. That type of behaviour was already observed in Fig.~\ref{fig:3}  but in the present case the borders between the different regimes become curvy, and the transition zones become wider. This means that for a variation of the parameters $\Lambda_{AB}$ and $\Lambda_B$ the state obtained varies more slowly and it is not so sensitive to interaction changes. Another interesting feature is the area around $\Lambda_B=-5$, which is the one explored in Fig.~\ref{fig:5}, because the difference between species $A$ and $B$ becomes larger. We can see how for $\Lambda_{AB}=0$, $\Lambda_A=4$ and $\Lambda_B=-5$ we have $A$ type bosons condensed (see Fig.~\ref{fig:6}(e)) and $B$ type bosons experimenting the transition. When $|\Lambda_{AB}|$ is increased  a transition zone to a condensate for the $A$ bosons appears. Notice that this transition is a consequence of the interspecies interaction. This is observed also for the entropy, the trace of the density matrix squared and the Schmidt gap (see Figs.~\ref{fig:7}(b), ~\ref{fig:7}(c) and ~\ref{fig:7}(d)). All these facts illustrate the effects of the interspecies interaction. Different behaviour of $A$ and $B$ is also reflected in Figs.~\ref{fig:7}(e) and~\ref{fig:7}(f) where $L-R$ entropies for $A$ and $B$ respectively are represented.
This entropy characterizes each kind of state for each species. In accordance to Fig.~\ref{fig:5}, for $\Lambda_B = -5$ in 7(e) depending on $\Lambda_{AB}$ we find seven differentiated regions going form  $\Lambda_{AB}<0$ to $\Lambda_{AB}>0$ corresponding to a localized peak on the left (Fig.~\ref{fig:5}(a)), a catlike (Fig.~\ref{fig:5}(b)), a wide single peak, a binomial-like state (Fig.~\ref{fig:5}(c)), a wide single peak, a catlike (Fig.~\ref{fig:5}(d)) and a localized peak on the right (Fig.~\ref{fig:5}(e)). For the same interaction parameters for type $B$ bosons there are only three regions in Fig.~\ref{fig:7}(f) corresponding to a localized peak on the left (Fig.~\ref{fig:5}(a)), a catlike (Figs.~\ref{fig:5}(b),~\ref{fig:5}(c) and~\ref{fig:5}(d)) and a localized peak on the right (Fig.~\ref{fig:5}(e)).

\subsection{Attractive intraspecies interaction $\Lambda_A$}

\begin{figure}[t]
\centering
\resizebox{\columnwidth}{!}{\input{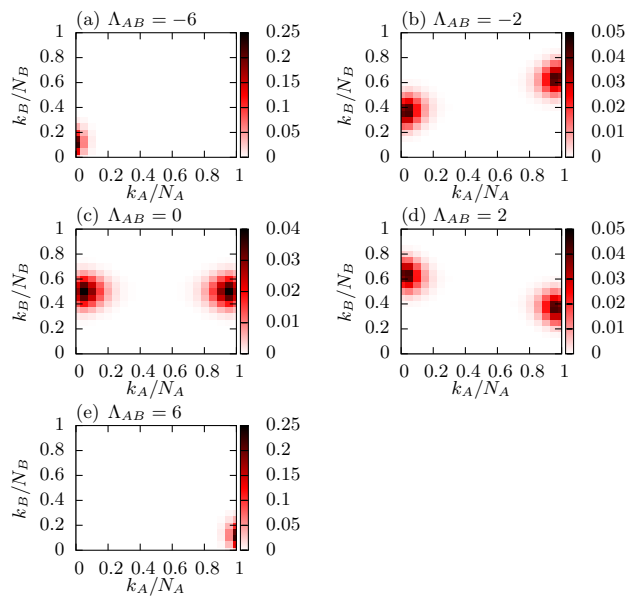}}
\caption{Spectral decomposition of the ground state, $|C_{k_A,k_B}|^2$, plotted for different values of $\Lambda_{AB}$. Here there is attraction between $A$ ($\Lambda_A<0$) and repulsion between $B$ type bosons ($\Lambda_B>0$). We observe the transition from (a), a regime dominated by the attraction between the different species to a regime (e) dominated by the repulsion between species $A$ and $B$, going through intermediate regimes in the panels (b) to (d). For all panels $\Lambda_A=-4$, $\Lambda_B=5$, $N_A=N_B=20$, $J=20$ and $\epsilon/J=10^{-10}$.}
\label{fig:8}
\end{figure}

\par
\begin{figure}[t]
\centering
\resizebox{\columnwidth}{!}{\input{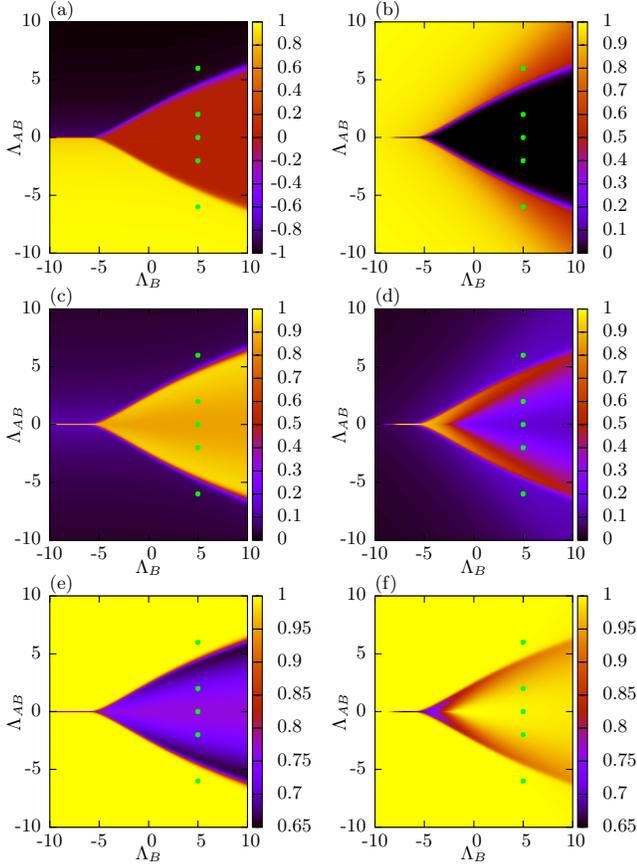}}
\caption{Properties of the ground state for $\Lambda_A<0$ and varying $\Lambda_B$ and $\Lambda_{AB}$. (a) Population imbalance $z_A$ and (b) $z_B$. Dispersion of population imbalance for each species, (c) $\sigma z_A$ and (d) $\sigma z_B$. Condensed fraction (e) $n^A_1$ and (f)  $n^B_1$. Green spots mark where the 
states of Fig.~\ref{fig:8} are found. For all panels $\Lambda_A=-4$, $N_A=N_B=20$, $J=20$ and $\epsilon/J=10^{-10}$.}
\label{fig:9}
\end{figure}

\begin{figure}[ht]
\centering
\resizebox{\columnwidth}{!}{\input{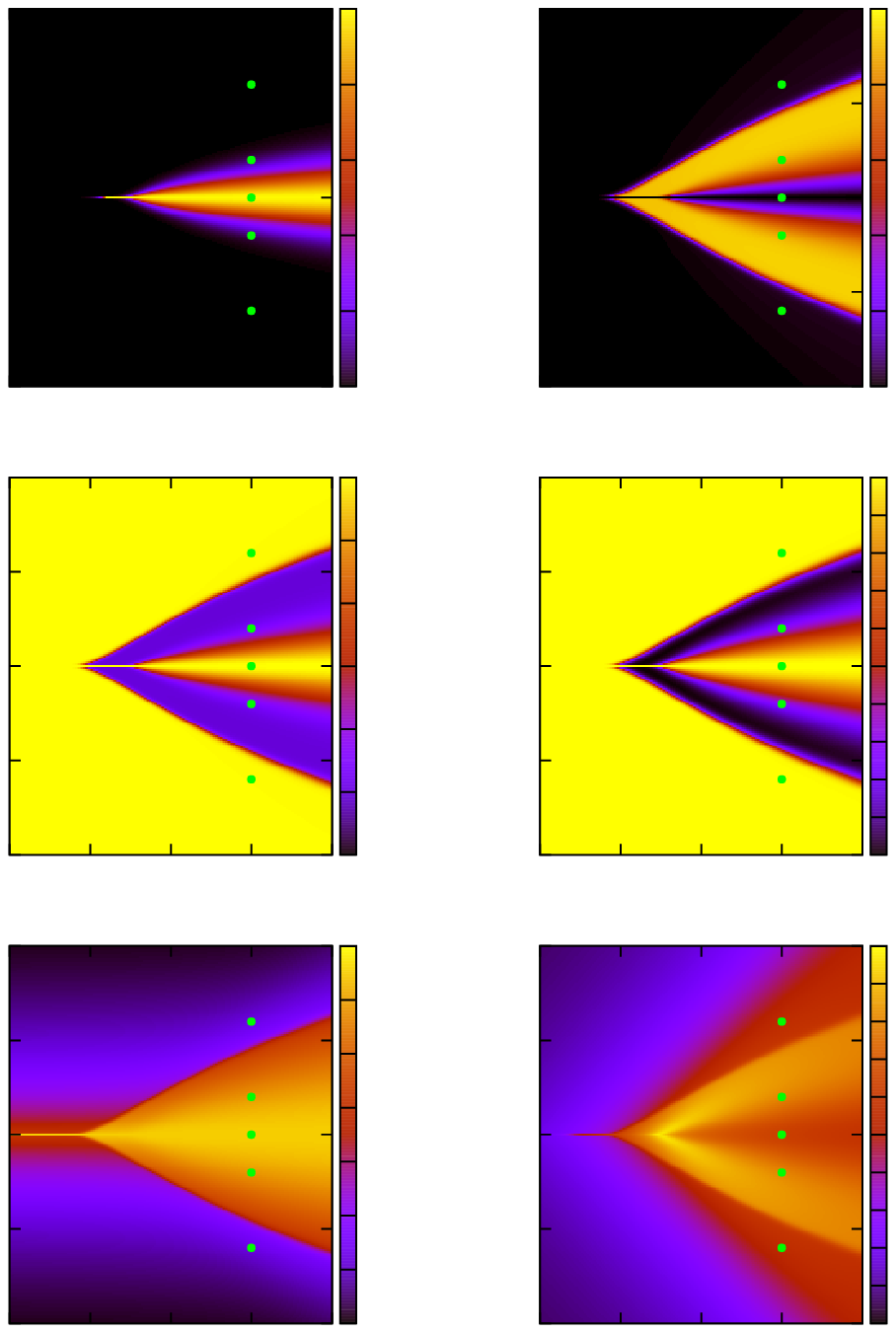}}
\caption{Properties of the ground state for $\Lambda_A<0$ and varying $\Lambda_B$ and $\Lambda_{AB}$. (a) Energy gap, $\Delta E_{1,0}$. (b) Von Neumann entropy for subsystem $A$($B$), $S_{A(B)}$ normalized to its maximum value. (c) Trace of the density matrix squared for subsystem $A$($B$), $P_{A(B)}$. (d) Schmidt gap for subsystem $A$($B$), $\Delta \lambda^{A(B)}$. (e) L$-$R von Neumann entropy for subsystem $A$, $S_{LR}^A$, and (f) for subsystem $B$, $S_{LR}^{B}$, both normalized to their maximum values. Green spots mark where the states of Fig.~\ref{fig:8} are found. For all panels $\Lambda_A=-4$, $N_A=N_B=20$, $J=20$ and $\epsilon/J=10^{-10}$.}
\label{fig:10}
\end{figure}

 In Figs.~\ref{fig:8}(a) and~\ref{fig:8}(e) we find similar type of states as those found before in Figs.~\ref{fig:5}(a) and~\ref{fig:5}(e). For $\Lambda_{AB}=0$ (see Fig.~\ref{fig:8}(c)), we have a catlike state for type $A$ bosons and a binomial-like state for type $B$ bosons. This situation is not different from the previous one in Fig.~\ref{fig:5} since it corresponds to an exchange of roles of $A$ and $B$ bosons. However, in Figs.~\ref{fig:9} and~\ref{fig:10} we can observe this situation from a different point of view because the variable parameter $\Lambda_B$ corresponds to the species which does the transition from binomial-like to highly localized state whereas for Figs.~\ref{fig:6} and~\ref{fig:7} bosons of type $B$ experimented the other transition.

\par On the one hand, for type $A$ bosons we see clearly three zones in Figs.~\ref{fig:9}(a), ~\ref{fig:9}(c) and~\ref{fig:9}(e). The top region is the one corresponding to have all bosons of this type on the right (see Fig.~\ref{fig:8}(e)) so its population imbalance is $-1$, the dispersion $0$ and there is condensation. The bottom region is similar to the top one but with $A$ bosons now confined on the left site (see Fig.~\ref{fig:8}(a)). The third region located on the right is the one corresponding to the catlike states for the $A$ species. On the other hand, for type $B$ bosons (see Figs.~\ref{fig:9}(b),~\ref{fig:9}(d) and~\ref{fig:9}(f)), the top and the bottom regions are the ones associated respectively with Fig.~\ref{fig:8}(e) and Fig.~\ref{fig:8}(a) and the right region is the transition where $B$ bosons pass from being confined in one side, to a catlike state, to a wide peak and finally a binomial-like state for $\Lambda_{AB}=0$.
\par In the present case, $\Lambda_A=-4$, we observe large regimes of $\Lambda_B$ and $\Lambda_{AB}$ for which there is degeneracy as we can observe in panel (a) of Fig.~\ref{fig:10}, where we report the energy gap between the ground and the first excited state. Moreover, the next three panels tell us that the presence of catlike states with entanglement, corresponding to the yellow zone in the entropy and purple zone in the trace of the density matrix squared, exists for a wide range of $\Lambda_{AB}$ (see Figs.~\ref{fig:10}(b),~\ref{fig:10}(c) and~\ref{fig:10}(d)).

\section{Different number of bosons}
\label{sec6}

In Sect.~\ref{sec4} and~\ref{sec5} we have reported results for the $N_A=N_B$ case. Now, we relax this condition and discuss the effect of having unequal populations.

\begin{figure}[t]
\centering
\resizebox{\columnwidth}{!}{\input{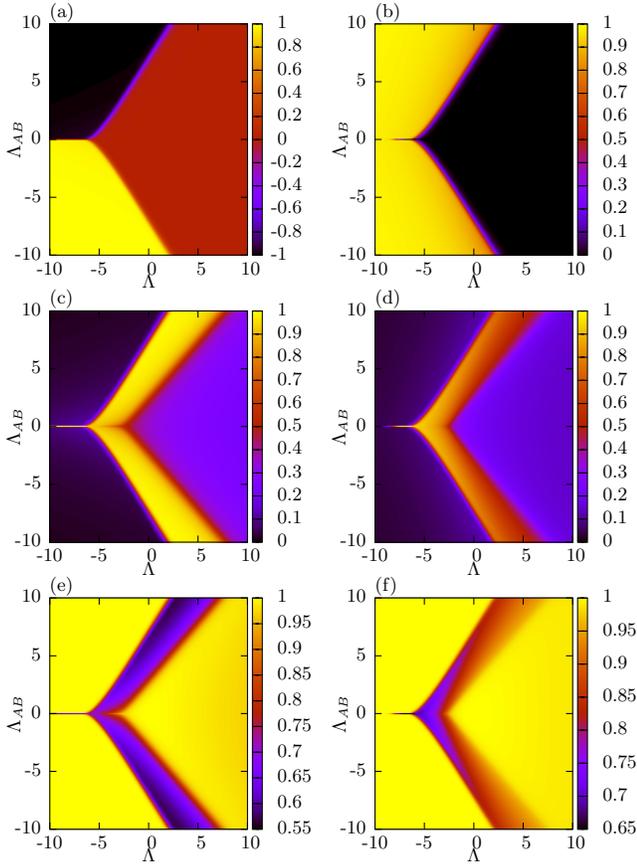}}
\caption{Properties of the ground state varying $\Lambda$ and $\Lambda_{AB}$. Population imbalance $z_A$ (a) and (b) $z_B$. Dispersion of population imbalance for each species, (c) $\sigma z_A$ and (d) $\sigma z_B$. Condensed fraction (e) $n^A_1$ and (f)  $n^B_1$. For all panels $N_A=8$, $N_B=20$, $J_A=8$, $J_B=20$ and $\epsilon/J_B=10^{-10}$.}
\label{fig:11}
\end{figure}

\begin{figure}[t]
\centering
\resizebox{\columnwidth}{!}{\input{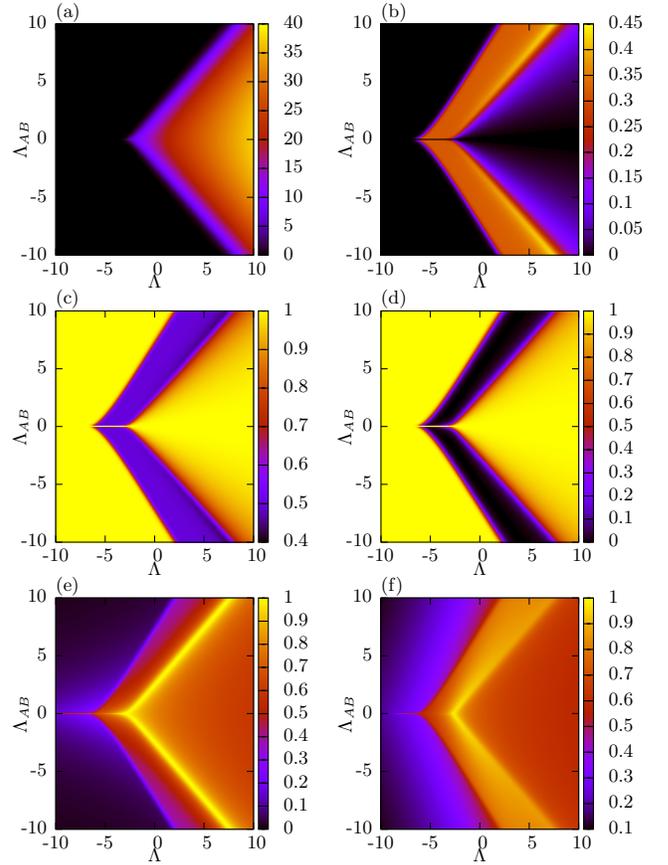}}
\caption{Properties of the ground state varying $\Lambda$ and $\Lambda_{AB}$. (a) Energy gap, $\Delta E_{1,0}$. (b) Von Neumann entropy for subsystem $A$($B$), $S_{A(B)}$ normalized to its maximum value. (c) Trace of the density matrix squared for subsystem $A$($B$), $P_{A(B)}$. (d) Schmidt gap for subsystem $A$($B$), $\Delta \lambda^{A(B)}$. (e) L$-$R von Neumann entropy for subsystem $A$, $S_{LR}^A$, and (f) for subsystem $B$, $S_{LR}^{B}$, both normalized to their maximum values. For all panels $N_A=8$, $N_B=20$, $J_A=8$, $J_B=20$ and $\epsilon/J_B=10^{-10}$.}
\label{fig:12}
\end{figure}

The effect of having different number of bosons is shown comparing Figs.~\ref{fig:11} and~\ref{fig:12} with Figs.~\ref{fig:3} and~\ref{fig:4}. In panels (a) and (b) of Fig.~\ref{fig:11} we can see that the population imbalance does not present differences from the corresponding ones of Fig.~\ref{fig:3} but the unequal number of bosons of each kind is reflected in the dispersion of the population imbalance (Figs.~\ref{fig:11}(c) and~\ref{fig:11}(d)) and in the condensed fraction (Figs.~\ref{fig:11}(e) and~\ref{fig:11}(f)). Now these two last quantities are not the same for both species as there were in Figs.~\ref{fig:3}(c) and~\ref{fig:3}(d) for same number of bosons. For species $A$ we observe a domination of yellow in the transition region in Fig.~\ref{fig:11}(c) and purple in Fig.~\ref{fig:11}(e) that indicates predominance of catlike state but for $B$, Figs.~\ref{fig:11}(d) and~\ref{fig:11}(f), the region is mainly red, which indicates a wider peak in the spectral decomposition for this species. The tiny region that is red for both species in these figures is when there is a major spread in the Fock space and corresponds to the maximum $A-B$ von Neumann entropy region in yellow in Fig.~\ref{fig:12}(b). $L-R$ von Neumann entropy for each species reflects also that for the same values of the inter and intraspecies interaction parameters we have different type of states depending on the number of particles. The observables studied indicate that if, for example, it is desired to have a $L-R$ catlike state as a ground state, we can achive it with this kind of mixture focusing on the species with less number of particles. In this way, the range of values of $\Lambda$ and $\Lambda_{AB}$ where this happens is wider than it was for same number of particles.

\section{Summary and Conclusions}
\label{sec7}

In this work we have discussed the ground state properties of a binary mixture of Bose-Einstein condensates in two spatial sites. The system has been described by means of a two-site two-component Bose-Hubbard Hamiltonian. Taking the same fixed number of particles for each component we have studied the properties of the ground state of the system in different interaction regimes, i.e. varying the intra and interspecies interactions and we have also studied the case of different number of particles of each species. The numerical tool used has been a direct diagonalization of the Hamiltonian which is feasible for the small number of particles considered, 20 for each component at most. In regimes where the interactions are much larger than the tunneling, the analytical ground state of the system can be obtained and can be used as a first approximation to the exact results. We have also considered a regime in which the interactions are of the order of the tunneling. In this case, sizeable quantum correlations are built in the system. First, we have discussed the symmetric case in which both components have the same intraspecies interaction and number of particles, finding quantum correlations arising as the interspecies interaction is tuned. Finally, we have also analyzed more general cases, first considering the case in which both species have a different intraspecies interaction and second the situation in which the number of particles of each type is not the same.

New type of states that cannot be found in a single-component condensate have been found and studied as the catlike ones that are of the interest for having entanglement between the two species bosons. We have discussed how the ground state can be characterized and how one can differentiate different qualitative ground states depending on $\Lambda_A$, $\Lambda_B$ and $\Lambda_{AB}$ using their properties. The population imbalance provides an average information that is complemented with the calculation of its dispersion. To determine the degree of entanglement, it has been shown that the von Neumann entropy gives detailed information about the state, clearly distinguishing among interesting correlated states.

Even if the phenomenology associated to a binary bosonic mixture in a double well is very broad we have tried to explore the most significant region of parameters with the hope that the detailed analysis reported in the paper can be helpful for the design and understanding of future experiments with mixtures, specially to identify where we can expect states with strong quantum correlations.
\vspace*{0.5cm}
\begin{acknowledgments}
The authors acknowledge financial support by grants 2014SGR-401 from Generalitat de Catalunya and FIS2014-54672-P from the MINECO (Spain). B.J.-D. is supported by the Ram\'{o}n y Cajal program.
\end{acknowledgments}

\end{document}